\documentclass[aps,pra,showpacs,twocolumn,groupedaddress,floatfix]{revtex4}
\usepackage{graphicx}
\usepackage{amsmath}
\usepackage{amssymb}
\begin{document}
\title{Numerical Analysis of Boosting Scheme for Scalable NMR
Quantum Computation}
\author{Akira SaiToh}
\email[Electronic address: ]{saitoh@qc.ee.es.osaka-u.ac.jp}
\author{Masahiro Kitagawa}
\email[Electronic address: ]{kitagawa.m@ee.es.osaka-u.ac.jp}
\affiliation{Graduate School of Engineering Science, Osaka University\\
1-3 Machikaneyama, Toyonaka, Osaka 560-8531, JAPAN}
\date{19 May 2003; last revised: 11 February 2005}

\begin{abstract}
Among initialization schemes for ensemble quantum computation beginning
at thermal equilibrium, the scheme proposed by Schulman and Vazirani 
[L. J. Schulman and U. V. Vazirani, in \textit{Proceedings of the 31st
 ACM Symposium on Theory of Computing (STOC'99)} 
(ACM Press, New York, 1999), pp. 322-329] is known for the simple 
quantum circuit to redistribute the biases (polarizations) of qubits and
small time complexity. However, our numerical simulation shows that the 
number of qubits initialized by the scheme is rather smaller than 
expected from the von Neumann entropy because of an increase in the sum 
of the binary entropies of individual qubits, which indicates a growth
in the total classical correlation. This result|namely, that there is
such a significant growth in the total binary entropy|disagrees with
that of their analysis.
\end{abstract}
\maketitle

\section{Introduction}
\begin{figure}
\scalebox{1.0}{\includegraphics{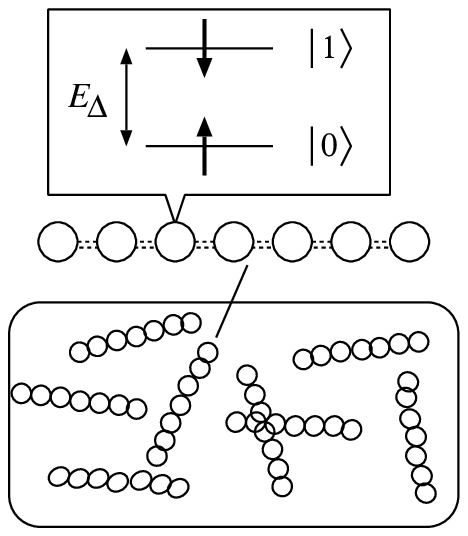}}
\caption{\label{fig1}A model of the NMR system for spin-1/2 nuclei.}
\end{figure}

Initialization is an indispensable stage in ensemble quantum computation
beginning at thermal equilibrium, especially in NMR computations
\cite{CFH97,GC97,CGK98,CVZ98,CMP98}. For a model of the NMR system, it 
is customary to consider a mixture of molecules with $n$ spins-1/2 under
a strong static magnetic field as shown in Fig.\ \ref{fig1} on the 
assumption of the independence among initial spins; local couplings among 
spins in a molecule are negligible compared with their Zeeman energies 
$E_\Delta$. The original bias (polarization) $\varepsilon$ of a spin at
temperature $T$ is given by 
\begin{equation}
 \varepsilon = \tanh\left(\frac{E_\Delta}{2k_BT}\right),
\end{equation}
where $k_B$ is Boltzmann's constant. This spin is in 
$|\hspace{-3pt}\uparrow\rangle$ with probability $(1+\varepsilon)/2$ 
and in $|\hspace{-3pt}\downarrow\rangle$ with probability 
$(1-\varepsilon)/2$. We regard the up spin as the bit of $0$ and the 
down spin as the bit of $1$. Quantum operations are performed in this 
$n$-qubit \cite{DE89,SC95} ensemble system. However, it contains $2^n$ 
\textit{component states} $|0\cdots 0\rangle,\ldots,|1\cdots 1\rangle$,
unless $T=0$; we have to extract the final state evolved from a
particular component state to obtain the outcome of a quantum
computation for any standard quantum algorithm. 
Suppose that the correct input is only the state $|0\rangle^{\otimes n}$.
While its original population is largest among all the component states, 
the product of probabilities of each bit being $0$ is small for large
$n$. The more bits we need, the larger biases are required to extract
the final state. As it is impossible to achieve $\varepsilon\simeq 1$ 
by cooling sample materials, algorithmic initializations 
\cite{CFH97,GC97,KCL98,SV98,SV99,RE99} are utilized to prepare an
initialized state: namely, a particular component state with an 
enhanced signal.

Standard quantum algorithms are designed to work with a pure-state
input. In contrast, there are a few quantum algorithms for the model
of deterministic quantum computation with a single pure qubit 
(DQC1) \cite{KL98} and similar models, including one for prime 
factorization \cite{PP99} (an extension of Shor's algorithm 
\cite{SH97}); those are designed to work with a highly mixed state, 
even with an input string that comprises a single pure qubit (namely, an
initialized qubit) and the rest qubits in a maximally mixed state. As 
the family of DQC1 models does not provide all the replacements of 
standard quantum algorithms, initialization of many qubits for ensemble 
quantum computers is still of importance.

There are two types of initialization methods: one is averaging ({\em e.g.},
effective pure states \cite{KCL98}); another is data compression.
The latter has an advantage in respect of scalability because the 
former needs resources growing exponentially in $n$ while the latter 
does not. The boosting scheme proposed by Schulman and Vazirani 
\cite{SV98,SV99} is one of the latter type of methods. Actually, 
their scheme is a sort of polarization transfer which has been commonly 
used and thoroughly studied in the realm of NMR spectroscopy as pointed 
out in Ref.\ \cite{CVS01}. There have been notable studies of the general
upper limit on a polarization enhancement in a polarization transfer 
({\em e.g.}, Refs.\ \cite{EBW87,SOL90,SOL91}). The principle of the 
boosting scheme is the bias-enhancing permutation as we will see 
in Sec.\ \ref{sec2}. In the scheme, a quantum circuit \cite{DE89} 
properly composed of basic bias-boosting circuits is used for 
redistributing the biases to generate the block of $l$ 
\textit{cold qubits} with biases greater than $\varepsilon_{cold}$ such 
that
\begin{equation}
 \mathrm{Prob}\{0_10_2\cdots 0_{l-1}0_l\}\gtrapprox
                       \left(\frac{1+\varepsilon_{cold}}{2}\right)^l
\end{equation}
is large enough to distinguish $|0\rangle^{\otimes l}$ among the signals
of $2^l$ component states. Their analyses \cite{SV98,SV99} resulted in 
that a compression down to almost the entropic limit is possible (in the
limit of large $n$) in $O(n\log n)$ time in the scheme to generate 
$n(1-o(1))-S$ initialized qubits each of which has bias 
$1-\exp(-n^{\Theta(1)})$ (which converges to $1$ as $n$ becomes large), 
where $S$ is the von Neumann entropy \cite{JVN27} of the ensemble 
\footnote{
For an original ensemble in a common NMR system, the sum of the binary
entropies of individual qubits is equal to the von Neumann entropy 
of the ensemble.}. However, this is incorrect unless the original biases
are in the neighborhood of $0$ or $1$
\footnote{
Their analyses of the scheme were based on the optimistic assumption that
the sum of the binary entropies is preserved during the initialization 
process of it.
}.
Irrespectively of $n$, the condition of original biases
$0<\varepsilon\lnapprox 1$ leads to the circumstance that the probability 
of a spin being up has classical correlations (classical dependences) 
with those of other spins within several steps after the beginning of 
the scheme. Owing to the correlations, $l$ can be much smaller than 
$n-S$. Especially for the practical setting of $n \le 10^3$, an accurate
evaluation of the scheme is of importance since a large mean bias such 
as $\langle\varepsilon\rangle = 0.4$ 
\footnote{An average polarization of 0.4 is a realistic
value. It was reported that an average proton polarization of 0.7 was 
attained in naphthalene by using photo-excited triplet electron spins 
\cite{TTT04,ITSOMYS00}. It was also recently reported that a pair of
almost pure qubits were achieved by a parahydrogen-induced polarization 
technique \cite{ABC04} and a quantum algorithm was implemented with the
qubits \cite{AJB04}.}
is required to obtain a sufficient number of initialized qubits \cite{BO01}.

In this Paper, we report a numerical analysis of the boosting scheme 
based on a molecular simulation. Several features of the scheme relating to 
the total binary entropy have been investigated by using the simulation 
with the aid of matrix calculations. It has revealed the fact that the 
number of qubits initialized by the scheme is rather smaller than
expected from the von Neumann entropy even for large $n$. This is not a 
surprise; it is quite possible according to the theory of macroscopic 
entropy described in Sec.\ \ref{sec3}.

\section{Boosting scheme}\label{sec2}
Enhancing the bias $\varepsilon_i$ of the $i$th spin is an operation
equivalent to increasing the population of a set of molecules whose 
$i$th spin is up [$\varepsilon_i=\varepsilon_i(t)$ where $t$ denotes 
time step]. This is a general conception of initializations for mixed 
states. As we consider a large number of molecules under a strong static 
magnetic field such as those in a common liquid- or solid-state NMR system, 
an $n$-qubit thermal equilibrium state can be written as the density matrix
\begin{equation}\label{eq3}
 \rho_{eq} = \rho^{eq}_1\otimes \rho^{eq}_2\otimes \cdots \otimes \rho^{eq}_{n-1}\otimes\rho^{eq}_n,
\end{equation}
where $\rho^{eq}_i$ is the density matrix representing the original state of 
the $i$th spin with the bias $\varepsilon_i(0)$:
\begin{equation}\label{eq4}
 \rho^{eq}_i = \begin{pmatrix}
\frac{1+\varepsilon_i(0)}{2}&0\\ 0&\frac{1-\varepsilon_i(0)}{2}
\end{pmatrix}.
\end{equation}
Let $c_k=c_k(t)$ denote the population of component state $|k\rangle$
at time step $t$.
Then it is also represented as
\begin{eqnarray}
 \rho_{eq} &=&\sum_{k=0}^{2^n-1}c_{k}(0)~|k\rangle\langle k|\nonumber\\
&=&\mathrm{diag}\{c_{0}(0), c_{1}(0), \ldots, c_{2^n-2}(0), c_{2^n-1}(0)\},
\end{eqnarray}
where $c_{k}(0)$ are the original populations. The bias of the $i$th
spin is enhanced by any unitary operation that permutes a component
state with a large population to some component state whose $i$th 
bit is $0$, since such an operation increases the population of a set of
component states whose $i$th bit is $0$:
\begin{equation}\label{eq6}
 P_i = \sum_{k|\text{$i$th bit is 0}} c_k.
\end{equation}
This population is identical to the probability that the $i$th bit is
$0$. Because a component state which consists of a larger number of 0's
and a smaller number of 1's has a larger population among the $2^n$ 
component states in most cases, any unitary transformation that maps a 
sufficient number of component states with more than $n/2$ 0's to those
with 0's in specific $l'$ bits enhances the biases of the $l'$ spins.
The initialization of $l\ge l'$ qubits can be realized by a combination 
of such transformations. When the transformation $U$ is only a product of 
permutations of component states, as it is nothing but the exchanges of 
values of $c_0,\ldots,c_{2^n-1}$, the evolved density matrix 
$\rho=U\rho_{eq}U^\dagger$ is still represented as
\begin{equation}
 \rho =\sum_{k=0}^{2^n-1} c_k |k\rangle\langle k|,
\end{equation}
although it is often impossible to write $\rho$ in the form of 
Eq. (\ref{eq3}).

\begin{figure}
\includegraphics{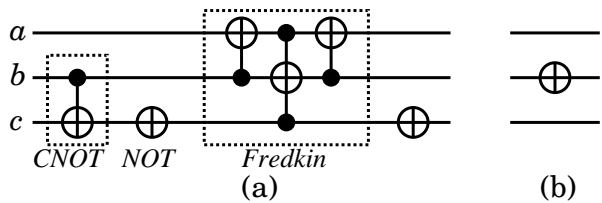}
\caption{\label{fig2}Basic quantum circuits for constructing an
initialization circuit in the boosting scheme. The CNOT gate performs
a bit flip (namely, a NOT operation) on the qubit $c$ when $b=1$. 
The Fredkin gate swaps the qubit $a$ with the qubit $b$ when $c=1$.}
\end{figure}

\begin{figure}
\begin{center}
\includegraphics[scale=1.0]{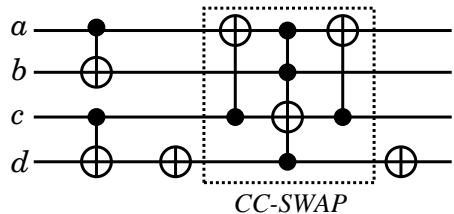}
\caption{\label{fig3}Another boosting circuit with a 4-qubit input. 
The CC-SWAP gate swaps $a$ with $c$ when $bd=11$.}
\end{center}
\end{figure}

In Schulman and Vazirani's boosting scheme (or simply called the boosting 
scheme), the total operation for an initialization can be composed of 
the 3-qubit basic circuits shown in Fig.\ \ref{fig2}
\footnote{An experiment of this boosting operation in
a three-spin system was reported in Ref.\ \cite{CVS01}.}
or the 4-qubit circuit shown in Fig.\ \ref{fig3}, 
which perform permutations of component states as mentioned above.
It should be noted that these circuits were not distinctly written in 
the original description of the boosting scheme. Either the 3-qubit 
circuits or the 4-qubit circuit can be a basic operation of the scheme.
If one interprets the original description as it is, one may have the 
4-qubit circuit. We, however, use the 3-qubit circuits rather than 
the 4-qubit circuit because the 3-qubit circuits are better with 
respect to the rate of initialization. We do not write details about 
it here, but we will describe a reason in Sec.\ \ref{sec4}. From now 
on, we will regard only the 3-qubit circuits as the basic circuits of the 
scheme.

The operation of the basic circuits is illustrated in the truth table 
of Table\ \ref{table1}. By the circuit of Fig.\ \ref{fig2}(a), 
a component state with a large population|{\em i.e.}, a component 
state in which two of the three qubits are 0|is mapped to a 
component state in which the first of the three qubits is 0. The 
probability that the first qubit $a$ is 0 after this basic operation is
\begin{eqnarray}
 \mathrm{Prob}\{a=0\}_{out}&=&
\mathrm{Prob}\{000\}_{in}+\mathrm{Prob}\{001\}_{in}\nonumber\\
&\ &+\mathrm{Prob}\{010\}_{in}+\mathrm{Prob}\{100\}_{in},
\end{eqnarray}
where, for given qubits $a$, $b$, and $c$ (from upper to lower in the 
circuit) with their binary values $x_a$, $x_b$, and $x_c$,
\begin{equation}
\mathrm{Prob}\{x_ax_bx_c\} = \sum_{k|abc=x_ax_bx_c}c_k.
\end{equation}
Suppose that no correlation exists among $P_a$,$P_b$, and $P_c$
and that the biases $\varepsilon_a$, $\varepsilon_b$, and $\varepsilon_c$
are positive, in the input. Then the bias of the qubit $a$ in the output
of the circuit is
\begin{equation}\label{eq10}
 \varepsilon'_a=\frac{1}{2}\left(
\varepsilon_a+\varepsilon_b+\varepsilon_c
-\varepsilon_a\varepsilon_b\varepsilon_c\right),
\end{equation}
and $\varepsilon'_a$ becomes larger than $\varepsilon_a$ under the condition:
\begin{equation}\label{ineq11}
 \varepsilon_a < 
          \frac{\varepsilon_b+\varepsilon_c}{1+\varepsilon_b\varepsilon_c}.
\end{equation}
Similar calculations give the biases for $b$ and $c$ in the same output:
\begin{eqnarray}\label{eq12}
  &\varepsilon'_b=\frac{1}{2}\left(
\varepsilon_a+\varepsilon_b-\varepsilon_c
+\varepsilon_a\varepsilon_b\varepsilon_c\right),\\\label{eq13}
  &\varepsilon'_c=\varepsilon_b\varepsilon_c.
\end{eqnarray}
After the circuit of Fig.\ \ref{fig2}(a), the bias of the qubit $b$ 
may be smaller than 0; the circuit of Fig.\ \ref{fig2}(b) is added 
to invert the bias in this case.
\begin{table}
\caption{\label{table1}Truth table for the basic circuits in
 Fig.\ \ref{fig2} for the boosting scheme. Here, the 
 input of (b) is the output of (a).}
\begin{ruledtabular}
\begin{tabular}{ccc}
\multicolumn{3}{c}{3-qubit boosting}\\\hline
Input&Output of (a)&Output of (b)\\\hline
000&000&010\\
001&001&011\\
010&011&001\\
011&100&110\\
100&010&000\\
101&101&111\\
110&111&101\\
111&110&100
\end{tabular}
\end{ruledtabular}
\end{table}
In the case where three spins are polarized uniformly with 
the bias $\varepsilon_{uni}$ in the input, Eq.\ (\ref{eq10}) leads to
\begin{equation} \label{eq14}
\varepsilon'_a=\frac{3}{2}\varepsilon_{uni}-\frac{1}{2}\varepsilon_{uni}^3.
\end{equation}
Thus, the basic circuits can be used for boosting the biases of target 
qubits. It is expected that one can achieve the component state 
$|0\rangle^{\otimes l}$ with a very large population simply by 
collecting initialized qubits after sufficient steps of application of 
these circuits to every subgroup which consists of three qubits that 
satisfy inequality\ (\ref{ineq11}) 
(although the initialization rate is another subject).
Equations\ (\ref{eq10}), (\ref{eq12}), and (\ref{eq13}) may be
used for a rough estimation of the behavior of a quantum circuit 
composed of the basic circuits.

However, the designing of a whole circuit in the boosting scheme is not 
so simple. At every step in the scheme, one recreates subgroups by 
picking up every three qubits from the $n$ qubits in the order of their 
biases, from larger to smaller, so that the condition of inequality\
(\ref{ineq11}) is satisfied in almost every subgroup. It is better to 
avoid applying the basic circuits to those in which the condition is not 
satisfied or undo it afterward. In order to recreate the subgroups, one 
has to forecast the biases of qubits correctly, which requires a more 
precise estimation owing to the following reason: After several steps 
in the scheme, the bias of a qubit is not independent of those of others, 
{\em i.e.},
\begin{equation}
 \mathrm{Prob}\{ij=00\}\not = P_iP_j
\end{equation}
for a pair of qubits $(i,j)$ in the $n$-qubit string. This is because
the ensemble after a CNOT operation on $(i,j)$ is not determined by
the biases $(\varepsilon_i,\varepsilon_j)$ alone but dependent on the 
populations $\{c_k\}$. The exception is only the case where the original 
biases are very close to $0$ or $1$. Therefore, the use of independent 
probabilities for representing spins in constructing the circuit induces
inaccuracy for the large size and/or large depth of the circuit. Each 
molecule has to be dealt with for correct forecasts of biases.

Moreover, even with the forecasts of biases, it is quite possible that 
the bias of the qubit $a$ is not boosted in some subgroups at almost 
every step except for several beginning steps in the scheme. This is due
to classical correlations among qubits. The condition of 
inequality\ (\ref{ineq11}) is not an accurate condition for the success of 
bias boosting in a subgroup of correlated qubits. It is nothing but a 
rough estimation of a correct condition. Thus, one ought to undo the 
operations that fail in enhancing the biases of target qubits so as to 
achieve better bias enhancements. The details of the circuit designing 
based on the forecasts and undo operations are described in 
Sec.\ \ref{sec5} and Appendix\ \ref{app3}.

Note that even the circuits that succeed in enhancing the biases alter 
the amount of correlation. Although the three qubits in each subgroup 
are nearly equally biased, this does not mean the binomial distribution, 
{\em i.e.}, it is not relevant to the independence of the qubits; their 
biases can be correlated [their individual probabilities given by Eq.\
(\ref{eq6}) can be dependent] to each other. The classical correlations 
affect the efficiency of initialization|this ``efficiency'' is not the 
efficiency in running time nor that in working space|as described in 
the next section.

\section{Classical correlations hidden in macroscopic binary entropies}
\label{sec3}
\begin{figure}
\begin{center}
\includegraphics[scale=0.65]{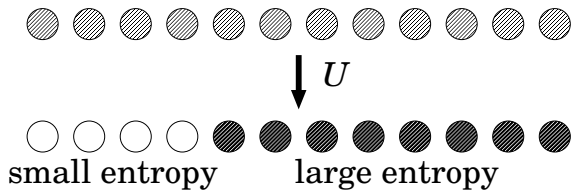}
\caption{\label{fig4}A unitary compression operation $U$ for
 redistributing the biases to create the block of qubits with small
 binary entropies and that with large binary entropies.}
\end{center}
\end{figure}

The initialization of an NMR quantum computer with a method that uses 
a data compression is equivalent to redistributing the binary 
entropies of qubits (the entropies of macroscopic bits),
$\{H_i=$
{\large$\frac{1+\varepsilon_i}{2}$}\hspace{2pt}$\log_2$\hspace{-2pt}
{\large$\frac{2}{1+\varepsilon_i}$}\hspace{2pt}$+$
{\large$\frac{1-\varepsilon_i}{2}$}\hspace{2pt}$\log_2$\hspace{-2pt}
{\large$\frac{2}{1-\varepsilon_i}$}$\}$,
to create two different blocks of qubits: one has a very small entropy 
close to $0$ and the other has a very large entropy as illustrated in 
Fig.\ \ref{fig4}.
The block of $l$ qubits with very small binary entropies is used as
$|0\rangle^{\otimes l}$. 
In this respect, an important entropy measure 
is the sum of the binary entropies of individual qubits: let the 
effective entropy 
\begin{equation}\label{eq16}
 S_e(t)= - \sum_{i=1}^{n} \left(
\frac{1+\varepsilon_i}{2}\log_2\frac{1+\varepsilon_i}{2} +
\frac{1-\varepsilon_i}{2}\log_2\frac{1-\varepsilon_i}{2}\right),
\end{equation}
where $\varepsilon_i=\varepsilon_i(t)$ is the bias of the $i$th qubit
at time step $t$. 
When the state of the whole $n$-qubit ensemble is represented as a 
diagonal density matrix, it is given by
\begin{equation}\label{eqsfbias}
 \varepsilon_i=-1+2\sum_{k|\text{$i$th bit is 0}}c_k.
\end{equation}
On the other hand, when there can be nonzero off-diagonal elements of 
the density matrix for the system, we have to consider the intrinsic bias 
defined below rather than the superficial bias that is calculated only 
from populations. The intrinsic bias of a qubit is a quantity obtained 
from a unitary diagonalization of the reduced density operator of the 
qubit. The reduced density operator of the $i$th qubit can be given by 
the partial trace
\begin{equation}
 \tilde\rho_i = \sum_{k=0}^{2^{n-1}-1}\langle\psi_k^i|\rho|\psi_k^i\rangle,
\end{equation}
where $\rho=\rho(t)$ is the density matrix for the whole ensemble; 
$\{|\psi_k^i\rangle\}=\{|\psi_0^i\rangle,\ldots,|\psi_{2^{n-1}-1}^i\rangle\}$
is a complete orthonormal system of the state space of the rest qubits, 
such as
\begin{equation}\begin{split}\{
 &|0_10_2\cdots 0_{i-1}0_{i+1}\cdots 0_{n-1}0_n\rangle,\\
 &|0_10_2\cdots 0_{i-1}0_{i+1}\cdots 0_{n-1}1_n\rangle,\\
 ~~~~~~&~~~~~~~~~~~~~~~\vdots\hspace{71pt},\\
 &|1_11_2\cdots 1_{i-1}1_{i+1}\cdots 1_{n-1}0_n\rangle,\\
 &|1_11_2\cdots 1_{i-1}1_{i+1}\cdots 1_{n-1}1_n\rangle
\}.
\end{split}\end{equation} 
Let ${}^d\tilde\rho_i$ be the matrix with the diagonal elements 
${}^dc_{i,0}$ and ${}^dc_{i,1}$ obtained from a unitary diagonalization of 
$\tilde\rho_i$ such that 
${}^dc_{i,0}\ge{}^dc_{i,1}\ge 0$. Then the intrinsic bias of the 
$i$th qubit is defined as 
\begin{equation}
 \hat\varepsilon_i= \frac{{}^dc_{i,0}-{}^dc_{i,1}}
{{}^dc_{i,0}+{}^dc_{i,1}}.
\end{equation}
This definition of intrinsic bias is valid in the sense that its value 
is independent of a macroscopic spin direction. The direction of the 
$i$th macroscopic bit $\tilde\rho_i$ can be modified even with only 
single-qubit rotation gates. If we restrict the subsequent quantum gates
to single-qubit rotations at a particular time step, 
$\frac{1+\hat\varepsilon_i}{2}$ is the maximum achievable probability of 
the $i$th qubit being $0$ after that step. Thus the intrinsic bias is a
plain measure of purity of a qubit. Since an initialization is to 
produce pure qubits, we regard $\hat\varepsilon_i$ as the actual bias of
the $i$th qubit denoted by $\varepsilon_i$ in the definition of $S_e$: 
namely, in Eq. (\ref{eq16}). When completely initialized qubits are 
required, $S_e$ is a lower bound for the length of the compressed string
at a particular time step in an initialization process; $n-S_e$ is an 
upper bound for the number of initialized qubits at that step 
\footnote{The optimal initialization for a time step $t$ is the one that 
achieves $\lfloor n-S_e(t)\rfloor$ qubits with biases of $1$ and 
$\lfloor S_e(t)\rfloor$ qubits with biases of $0$; there may be the rest
one qubit, which is included into the compressed string.}. 
$S_e$ varies step by step during the initialization.

Now we will see an upper bound for the number of initialized qubits 
under a realistic condition. Suppose that we do not need completely 
initialized qubits. Then initialized qubits are allowed to have 
small binary entropies ($H_j$) whose average value
($\frac{1}{l}\sum_{j=1}^lH_j$) is equal to or less than some real number
$\alpha$ ($0\le \alpha\ll 1$).
The total binary entropy of $l$ initialized qubits is equal to or less
than $\alpha l$. The block of rest $n-l$ qubits (namely, the compressed
string) therefore has the total binary entropy $E\ge S_e-\alpha l$. 
Hence the following inequality holds:
\begin{equation}
 n-l\ge E \ge S_e-\alpha l.
\end{equation}
This leads to
\begin{equation}
l\le\frac{1}{1-\alpha}(n-S_e).
\end{equation}
Setting $\beta=\frac{\alpha}{1-\alpha}$, we have
\begin{equation}\label{eqlupper}
 l\le (1+\beta)(n-S_e) \le n(1-S_e/n+\beta).
\end{equation}

In an initialization, even if there is no need of completely initialized
qubits, one picks up $l$ cold qubits at the terminal step of it in order
to achieve $\mathrm{Prob}\{0_1\cdots 0_l\}\ge c$ ($0\ll c\le 1$ is a 
real number). The value of $c$ is chosen by one's need. 
Accordingly, a limitation of the value of $\alpha$ can be estimated as 
follows: Suppose that the density matrix for a state after 
initialization has no nonzero off-diagonal element. Let us assume for
a moment that one obtains $l$ initialized qubits with independent 
biases and the rest qubits with biases that may be correlated. 
If the joint probability of the initialized qubits being 
zeros is $\mathrm{Prob}\{0_1\cdots 0_l\}\ge b$ ($0\le b\le 1$), then 
the binary entropies $H_j$ for the qubits satisfy the inequality 
$\sum_{j=1}^{l}H_j\le -l\left[b^{1/l}\log_2 b^{1/l}+
(1-b^{1/l})\log_2(1-b^{1/l})\right]$,
which is easy to prove by mathematical induction. From the 
contraposition, if 
$ \sum_{j=1}^{l}H_j> -l\left[b^{1/l}\log_2 b^{1/l}+
(1-b^{1/l})\log_2(1-b^{1/l})\right]$, then
$\mathrm{Prob}\{0_1\cdots 0_l\}< b$. Therefore, on the present
assumption that the biases of initialized qubits are independent, a
necessary condition for achieving $\mathrm{Prob}\{0_1\cdots 0_l\}\ge c$ 
is that
\begin{equation}\label{ineq24}
 \alpha\le -c^{1/l}\log_2 c^{1/l}-(1-c^{1/l})\log_2(1-c^{1/l}).
\end{equation}
Hence on the present assumption, $\alpha$ tends to $0$ as $l$
increases, and so does $\beta$. Possible values of $\alpha$ and $\beta$ 
are estimated with inequality\ (\ref{ineq24}) by specifying a value of $c$ 
and assuming that $l$ is in a realistic range. For example, when one 
needs $c=0.99$ and assumes $l\ge 16$, the values of $\alpha$ and 
$\beta$ are estimated to be at most $7.59\times10^{-3}$ and 
$7.64\times10^{-3}$ respectively. Indeed, it is, however, possible 
that $\alpha$ does not tend to $0$ with increasing $l$ to achieve 
$\mathrm{Prob}\{0_1\cdots 0_l\}\ge c$ when there are correlations among 
initialized qubits, but even for such qubits, of course the condition 
$\alpha\le -c\log_2 c-(1-c)\log_2(1-c)$ is necessary. Moreover, in the
case where one needs the bias of an initialized qubit to converge to 
$1$ as $n$ grows|{\em e.g.} $\varepsilon_{cold} = 1-\exp(-n^x)$ 
(constant $x>0$)|$\alpha$ and $\beta$ tend to $0$ as $n$ 
increases|{\em i.e.}, $\beta=o(1)$ in this case. As we have seen, 
inequality\ (\ref{eqlupper}) gives an upper bound for the number of 
initialized qubits at a particular time step for a realistic 
initialization. In other words, $S_e-\beta(n-S_e)$ is a lower bound for 
the length of the compressed string at that step. $\beta$ is a small 
number, which is estimated to be less than $10^{-2}$ in most of 
realistic initializations to make several bytes almost pure.

Generally, it is believed to be possible to almost faithfully compress 
the string of an $n$-qubit ensemble, represented by density matrix 
$\sigma$, down to the entropic limit with the rate $S(\sigma)/n$ for 
large $n$ in all kinds of quantum computer including NMR computers 
(as long as $n$ can be large) because of the analogy to the quantum 
noiseless coding \cite{SC95}, where
\begin{equation}
S(\sigma)=-\mathrm{Tr}(\sigma\log_2\sigma)
\end{equation}
is its von Neumann entropy. It is desirable, and believed to be
possible, to generate a quantum circuit to accomplish it within the 
depth of the circuit increasing polynomially in $n$. Indeed any 
unitary transformation of a density matrix conserves its von Neumann 
entropy, but it would be probably impossible for a compression scheme 
which generates a quantum circuit according only to macroscopic 
quantities of individual qubits, {\em e.g.}, biases, to do such an 
ideal compression for an arbitrary input state. For the initialization 
of NMR quantum computers, this is somewhat clear when we rewrite $S_e$ 
by using the reduced density operators $\{\tilde\rho_i\}$ of individual 
qubits. Regarding $\hat\varepsilon_i$ as the actual bias of the $i$th 
qubit, we have $H_i = S(\tilde\rho_i)$. Consequently, $S_e$ is 
rewritten as
\begin{equation}\label{eqSebyReduced}
S_e=\sum_{i=1}^{n}S({\tilde\rho_i}). 
\end{equation}
From this notation, it is clear that 
\begin{equation}\label{eq22}
 S_e-S=S(\rho~\|~{\tilde\rho_1}\otimes\cdots\otimes{\tilde\rho_n}),
\end{equation}
where $S=S(\rho)$, and 
$S(\rho~\|~{\tilde\rho_1}\otimes\cdots\otimes{\tilde\rho_n})$ 
is the Umegaki relative entropy \cite{UME62} of $\rho$ with respect to
${\tilde\rho_1}\otimes\cdots\otimes{\tilde\rho_n}$.
In general, the Umegaki relative entropy between two density operators 
$\sigma_1$ and $\sigma_2$ is defined as
\begin{equation}
 S(\sigma_1\|\sigma_2)=\mathrm{Tr}\left(\sigma_1\log_2\frac{\sigma_1}{\sigma_2}\right).
\end{equation}
It is a measure of the distance (or dissimilarity) between the two operators.
On the right-hand side of Eq.\ (\ref{eq22}), $\rho$ represents the whole
ensemble while ${\tilde\rho_1}\otimes\cdots\otimes{\tilde\rho_n}$ is a 
tensor product in which correlations are freed 
({\em i.e.}, correlations among ${\tilde\rho_1},\ldots,
{\tilde\rho_n}$ are obliterated in taking the tensor product).
Hence $S_e-S$ is a measure of total correlation \cite{HV01} of $\rho$: 
namely, a measure of total correlation of each qubit to the other
qubits. Obviously, $S_e$ can be much larger than $S$ when there is a 
growth in the total correlation during the initialization. There is no 
known way to avoid the creation and growth of correlations among biases 
when a data-compression circuit is designed by using solely the values of 
biases at every step instead of those of commonly-used quantities 
such as the (approximate) populations of possible bit sequences
\cite{HAM80}, the (approximate) length and frequency of patterns in 
microscopic strings, etc.

Let us consider the boosting scheme, in which every operation is a 
permutation of diagonal elements in a diagonal density matrix which 
is originally $\rho_{eq}$ as we have seen in the previous section. 
Then, clearly, there is no entanglement during and after a bias boosting
process; $S_e-S$ is a measure of total classical correlation hidden in 
the macroscopic string (the string which consists of $n$ macroscopic bits 
${\tilde\rho_1},\ldots,{\tilde\rho_n}$). Thus, in the boosting scheme, 
an increase in $S_e$ measures a growth in the total classical correlation.

In addition, the maximal efficiency of initialization of the 
macroscopic string is defined by
\begin{equation}
r_e = \frac{n-S_e^{end}}{n-S},
\end{equation} 
where $S_e^{end}$ is the value of $S_e$ after the application of an
initialization scheme. This is the ratio of an upper bound for the 
number of completely initialized qubits that can be achieved by the 
scheme to the one that is expected from the von Neumann entropy. 
Similarly, the maximal efficiency of compression of the macroscopic 
string is defined by 
\begin{equation}
r_c=S/S_e^{end}.
\end{equation}
One can evaluate the performance of the scheme by calculating these 
efficiencies.

\section{Growth in the amount of correlation for a basic boosting step}\label{sec4}
As we have seen in Sec.\ \ref{sec3}, an increase in the amount of 
correlation must be suppressed to achieve a high initialization rate.
In this section, we therefore evaluate the amount of classical correlation
generated by a single step of basic boosting circuits. 

\begin{figure}
\includegraphics[scale=0.6]{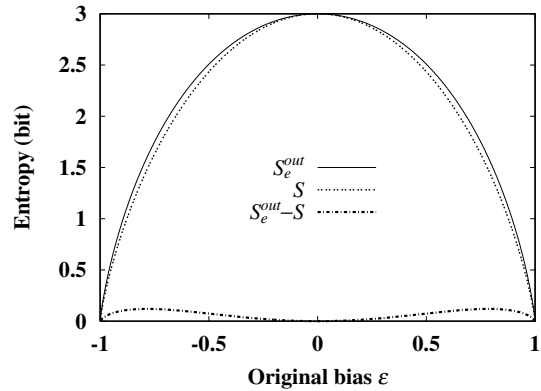}
\caption{\label{fig5} Plots of $S_e^{out}$, $S$, and $S_e^{out}-S$ for 
$3$ qubits as functions of $\varepsilon$, the uniform original bias of 
the qubits. $S_e^{out}$ is the effective entropy of the output state 
after one step of the basic circuit shown in Fig.\ \ref{fig2}(a) and 
$S$ is the von Neumann entropy of the system.}
\end{figure}

Suppose that there are three qubits $a$, $b$, and $c$ with 
uniform bias $\varepsilon$ at thermal equilibrium. For this 
initial state, the effective entropy defined by Eq.\ (\ref{eq16}) is 
equal to the von Neumann entropy $S$. Now we apply the 3-qubit circuit 
of Fig.\ \ref{fig2}(a) to the qubits. The biases of the qubits become 
$\frac{3\varepsilon-\varepsilon^3}{2}$,
$\frac{\varepsilon+\varepsilon^3}{2}$, and $\varepsilon^2$,
respectively. By this operation, classical correlations are induced,
and the effective entropy of the output state, $S_e^{out}$, is greater 
than $S$ as shown in Fig.\ \ref{fig5}. The discrepancy between 
$S_e^{out}$ and $S$ is rather serious for the highly but not completely 
polarized qubits because $S_e^{out}-S$ is fairly large in despite of the
small values of $S$.

\begin{figure}
\includegraphics[scale=0.6]{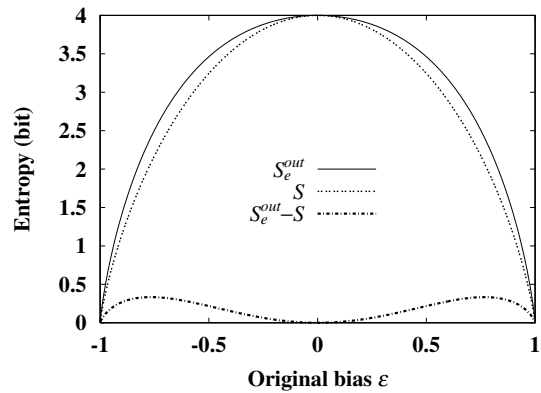}
\caption{\label{fig6} Plots of $S_e^{out}$, $S$, and $S_e^{out}-S$ 
for 4 qubits as functions of $\varepsilon$. $S_e^{out}$ is the 
effective entropy of them after one step of the circuit shown in 
Fig.\ \ref{fig3}.}
\end{figure}

Similarly, we can evaluate the amount of classical correlation generated
by the 4-qubit circuit shown in Fig.\ \ref{fig3}. When there are 4 
input qubits $a$, $b$, $c$, and $d$ with uniform bias $\varepsilon$ 
at thermal equilibrium, the biases in the output of it are 
$\frac{3\varepsilon-\varepsilon^3}{2}$,
$\varepsilon^2$, $\frac{\varepsilon+\varepsilon^3}{2}$, and 
$\varepsilon^2$, respectively. Thus the effect of bias enhancement 
looks almost the same as that of the 3-qubit basic circuits.
It, however, induces a larger amount of correlation in comparison with 
the 3-qubit basic circuits. The discrepancy between the effective
entropy of the output, $S_e^{out}$, and the von Neumann entropy $S$ 
is shown in Fig.\ \ref{fig6}. 

\begin{figure}
\begin{center}
 \includegraphics[scale=0.6]{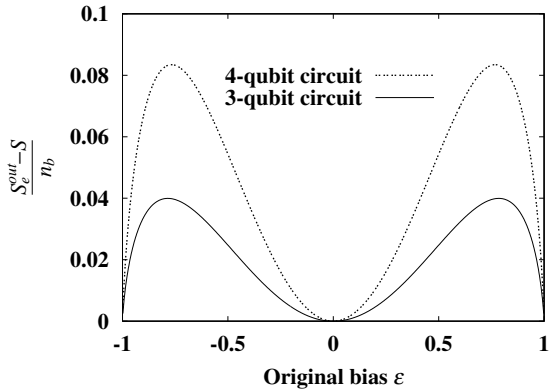}
\caption{\label{fig7}Plots of $(S_e^{out}-S)/{n_b}$ for the 3-qubit 
circuit ($n_b=3$) and the 4-qubit circuit ($n_b=4$) as functions of 
the original bias $\varepsilon$.}
\end{center}
\end{figure}

When one compares the mean discrepancy $(S_e^{out}-S)/{n_b}$ 
($n_b=3,4$ is the number of wires) for the 3-qubit circuit with that 
for the 4-qubit circuit, as shown in Fig.\ \ref{fig7}, it is clear 
that the increase in the amount of correlation in one time step is so 
large that the 4-qubit circuit should not be used as a basic boosting 
operation. This is the reason why we have chosen the 3-qubit circuits.

\section{Automatic circuit generation}\label{sec5}
\begin{figure}[bt]
\begin{center}
\includegraphics[scale=0.6]{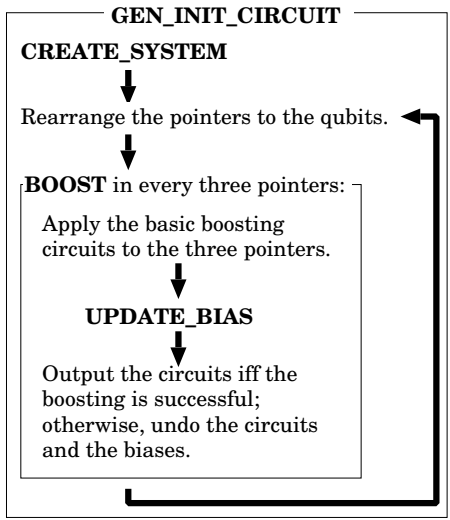}
\caption{\label{fig8}The brief flowing chart of the simulation
algorithm of Program\ 1 to generate an initialization circuit on the basis of
the boosting scheme. We show the program in Appendix\ \ref{app3}.
}
\end{center}
\end{figure}

A molecular simulation is applied for designing a quantum initialization
circuit in the boosting scheme on conventional computers. It is 
implemented as a circuit composer for the scheme, offering the forecasts
of biases during the initialization. The details of the algorithm for 
the designing are described in Program\ 1 of Appendix\ \ref{app3}. In 
this section, instead of the long program, the brief flowing chart of it
is shown in Fig.\ \ref{fig8}. It utilizes a virtual molecular system to 
mimic an $n$-qubit ensemble as illustrated in Fig.\ \ref{fig1}. As we 
continue to use the same basis 
$\{|0\cdots0\rangle,\ldots,|1\cdots1\rangle\}$, superpositions of
multiple computational basis vectors are not produced by the boosting 
operations. Hence a molecule is represented as an array of $n$ binary 
digits in the simulation. We consider $N$ molecules that consist of $n$ 
spins with bias $\varepsilon_i$ in the $i$th spin ($N\gg n$).

The outline of Program\ 1 is as follows: 
A table of pointers to qubits is used instead of direct handling
of them. We consider the subgroups which consist of every three pointers
from the upper side to the lower side in the table, although those are not
explicitly written in the program. Let $\text{frand()}$ be a function 
which returns a pseudo random real number between $0$ and $1$. This is 
called $n\hspace{-2pt}\cdot\hspace{-2pt}N$ times in the procedure 
\textbf{CREATE\_SYSTEM} to set up the virtual molecular system. 
At each call of $\text{frand()}$, the bit representing the $i$th spin 
in the $m$th molecule is set to $0$ when 
$\text{frand()} < (1+\varepsilon_i(0))/2$ and set to $1$ otherwise. The 
desired original biases $\varepsilon_1(0),\ldots,\varepsilon_n(0)$ are 
prepared in this way. Then the pointers are arranged in the order of 
the biases, from the larger to the smaller, in the table (from the 
upper side to the lower side). The main part of this simulation after 
the setup is an iteration in the procedure \textbf{GEN\_INIT\_CIRCUIT}. 
This iteration contains the following operations: 
Firstly, \textbf{BOOST} is applied to every subgroup mentioned above.
\textbf{BOOST} is the boosting operation for three qubits $a$, $b$, and
$c$. We apply the circuit of Fig.\ \ref{fig2}(a) to them. When the bias
of $b$ becomes negative, the circuit of Fig.\ \ref{fig2}(b) is also applied.
The boosting circuits applied to the three qubits are written out 
if the bias of $a$ is increased by them; otherwise, we undo the boosting.
The biases after the boosting are calculated by using
\textbf{UPDATE\_BIAS} in each call of \textbf{BOOST}. Secondly, we rearrange 
the table so that the pointers to qubits with larger biases appear 
in the upper side of it and those with smaller in the lower side. The 
iteration of above operations is terminated when cold qubits are not 
newly produced in the last $s_t$ steps or when the depth of generated  
circuits reaches a specified terminal depth of circuits.
After this iteration, qubits with very large biases are pointed from 
pointers located near the top of the table; the component state 
$|0\rangle^{\otimes l}$ with a very large population is achieved by 
collecting them.

This algorithm assumes that a CNOT operation can be used with any two
qubits in the system. When this is not guaranteed for dispersed
qubits, SWAP operations are used to gather them into one location 
in the molecule. This change is easily realized by arranging the qubits 
with SWAP gates in \textbf{GEN\_INIT\_CIRCUIT} instead of arranging 
only the pointers.

The complexity of the algorithm is $\mathrm{\Theta}(n\cdot N)$ in space 
and $O(n^2\cdot N)$ in time in conventional computers. For a practical
use of the simulation for $n \le 10^4$, it would be proper to set 
$N\simeq 10^4n$ and $s_t=5+\lfloor 0.1n\rfloor$. Our simulation can
be used for generating the circuits up to $10^4$ qubits for arbitrary 
biases within $1$ terabit of memory space.

In addition, the present algorithm is equivalent to the original one 
proposed by Schulman and Vazirani in Ref.\ \cite{SV99}. In the 
iteration of the original algorithm, at each step, qubits are 
rearranged so that qubits with large biases appear first followed by 
qubits with relatively small biases, discarding those with very low 
biases. Note that we do not need to discard qubits with very low biases 
actually. They are dispersed far away from those with large biases by 
the rearrangement automatically. In the present algorithm, in the 
iteration, the pointers to qubits are rearranged in the order of the 
biases. Therefore, it is equivalent to the original one.

\subsubsection*{Example A}\label{example2}
\begin{figure}[hpbt]
\includegraphics[scale=0.54]{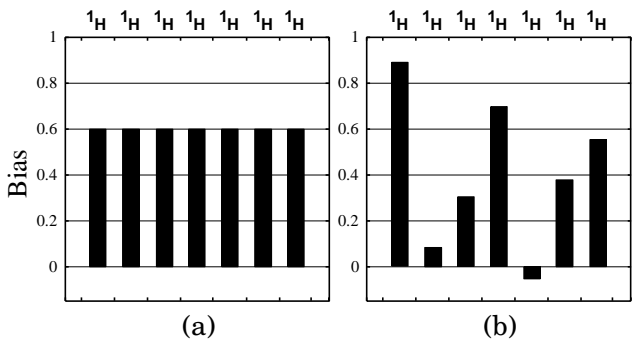}
\caption{\label{fig9}
The result from a matrix calculation for the first example. 
(a) The original biases of the 7-spin system.
(b) Redistributed biases after the operations of Table\ \ref{table2}.
One of the $\mathrm{{}^{1}H}$ spin biases is boosted under the preset
low temperature such that the original bias of the hydrogen is $0.6$,
in this matrix calculation.
}
\end{figure}

An example \footnote{The matrix calculation in these two examples
was done with an interpreter-type quantum circuit simulator
(\textit{http://www.qc.ee.es.osaka-u.ac.jp/\~{}saitoh/silqcs/}) that internally
uses the GAMMA C++ library (\textit{http://gamma.magnet.fsu.edu/})
created by Smith {\em et al.} \cite{SLME94}.}
of the output circuit from the program of Program\ 1 is shown in 
Table\ \ref{table2}. It is generated for the 7-pin system,
$\mathrm{{}^1H\text{-}{}^1H\text{-}{}^1H\text{-}{}^1H\text{-}{}^1H\text{-}{}^1H\text{-}{}^1H}$
with the uniform bias $0.6$. Figure\ \ref{fig9} shows that, according to
the numerical matrix calculation result, the bias of the first spin is 
enhanced as it is designed to. Although it appears to be possible to 
enhance the bias further at a glance, it is impossible for the boosting
scheme owing to strong classical correlations among the qubits. An NMR 
experiment of this bias boosting will be hopefully possible in future.
\begin{table}[hbpt]
\caption{\label{table2}
The initialization circuit generated for the spin system, 
$\mathrm{{}^1H\text{-}{}^1H\text{-}{}^1H\text{-}{}^1H\text{-}{}^1H\text{-}{}^1H\text{-}{}^1H}$.
The spins are labeled $1,\ldots,7$, from left to right, respectively.
The transformations in this table are executed from left to right and 
upper to lower. X, CN, and Fr read NOT gate, controlled-NOT gate, and 
Fredkin gate (controlled-SWAP gate), respectively, followed by numbers 
in parentheses indicating which qubits are handled. In CN($a$,$b$), the 
control bit is $a$ and the target bit is $b$; in Fr($a$ $b$, $c$), the 
control bit is $c$ and the target bits are $a$ and $b$.}
{
\begin{ruledtabular}
\begin{tabular}{l}
CN(2,3);X(3);Fr(1 2, 3);X(3);\\
CN(5,6);X(6);Fr(4 5, 6);X(6);\\
CN(2,6);X(6);Fr(7 2, 6);X(6);\\
CN(2,3);X(3);Fr(6 2, 3);X(3);\\
CN(6,5);X(5);Fr(3 6, 5);X(5);\\
CN(5,3);X(3);Fr(6 5, 3);X(3);\\
CN(4,7);X(7);Fr(1 4, 7);X(7);
\end{tabular}
\end{ruledtabular}
}
\end{table}
\subsubsection*{Example B}
\begin{figure}[hpbt]
\includegraphics[scale=0.54]{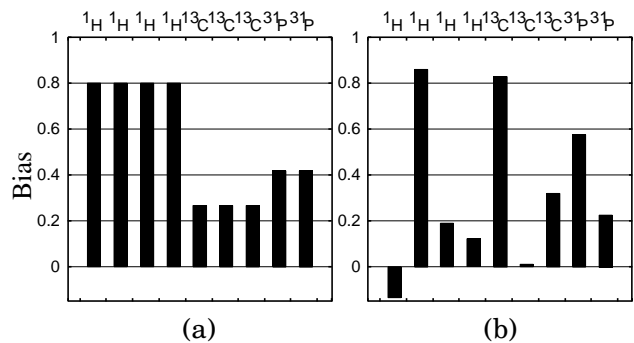}
\caption{\label{fig10}
The result from a matrix calculation for the second example.
(a) The original biases of the 9-spin system.
(b) Redistributed biases after the operations of Table\ \ref{table3}.
One of the $\mathrm{{}^{13}C}$ spin biases is boosted under the preset 
low temperature such that the original bias of the hydrogen is $0.8$,
in this matrix calculation.
}
\end{figure}

The second example is an output circuit for qubits with initially 
nonuniform biases. The circuit shown in Table\ \ref{table3} is generated 
for the 9-spin system,
$\mathrm{{}^1H\text{-}{}^1H\text{-}{}^1H\text{-}{}^1H\text{-}{}^{13}{C}\text{-}{}^{13}{C}\text{-}{}^{13}{C}\text{-}{}^{31}{P}\text{-}{}^{31}{P}}$.
We used a modified program to boost mainly the bias of the 5th spin.
The numerical matrix calculation of the boosting with the circuit 
results in an increase in the bias of one of the $\mathrm{{}^{13}C}$ 
spins as expected from the molecular simulation (Fig.\ \ref{fig10}). 
Of course, the effect of the boosting is too small to obtain an 
initialized qubit in such a small molecule. 
\begin{table}[hbpt]
\caption{\label{table3}
The initialization circuit generated for the spin system, 
$\mathrm{{}^1H\text{-}{}^1H\text{-}{}^1H\text{-}{}^1H\text{-}{}^{13}{C}\text{-}{}^{13}{C}\text{-}{}^{13}{C}\text{-}{}^{31}{P}\text{-}{}^{31}{P}}$.
The spins are labeled $1,\ldots,9$, from left to right, respectively.
The notation of the operations in this table is the same as that in 
Table\ \ref{table2}.}
{
\begin{ruledtabular}
\begin{tabular}{l}
CN(6,7);X(7);Fr(5 6, 7);X(7);\\
CN(9,2);X(2);Fr(8 9, 2);X(2);\\
CN(6,9);X(9);Fr(7 6, 9);X(9);\\
CN(1,4);X(4);Fr(2 1, 4);X(4);\\
CN(6,1);X(1);Fr(9 6, 1);X(1);\\
CN(4,3);X(3);Fr(7 4, 3);X(3);\\
CN(4,3);X(3);Fr(9 4, 3);X(3);X(4);\\
CN(7,8);X(8);Fr(5 7, 8);X(8);
\end{tabular}
\end{ruledtabular}
}
\end{table}

\section{Results from simulation}\label{sec7}
The simulation of Program\ 1 has been conducted under several different 
conditions of the original biases and the number of qubits, with the 
number of virtual molecules $N=5.0\times 10^6$. This value of $N$ is 
large enough to ensure the reliability of the simulation as we show 
in Appendix\ \ref{app1}. $\varepsilon_{cold}$ was set to 
$2\times0.9^{1/\lceil n-S\rceil}-1$ (this is larger than $0.99$ when 
$n-S\ge 22$) in Program\ 1. ($S$ is the von Neumann entropy of a 
simulated system.) As described in Sec.\ \ref{sec3}, we propose to 
compute the sum of the binary entropies of individual qubits during 
the simulation
because it is a lower bound for the length of the compressed string at
a particular time step $t$, at least approximately, in the boosting scheme. 
Recall that the effective entropy is thus the total binary entropy 
given by 
\begin{equation}
 S_e(t)=\sum_{i=1}^{n}\left(\frac{1+\varepsilon_i}{2} \log_2 
\frac{2}{1+\varepsilon_i} + \frac{1-\varepsilon_i}{2} \log_2 
\frac{2}{1-\varepsilon_i}\right)\tag{\ref{eq16}'}
\end{equation}
($\varepsilon_i=\varepsilon_i(t)$) and that $S_e$ can be increased by 
the growth of classical correlations among qubits in the scheme. At 
the last step in the simulation, we ignored the preset value of 
$\varepsilon_{cold}$ and picked up $l$ cold qubits so that the 
condition of $\mathrm{Prob}\{0_1\cdots 0_l\}>0.9$ was satisfied; 
although this does not affect the initialization process itself, nor 
does it affect the time evolution of the effective entropy. Here we
adopt the circuit of Fig.\ \ref{fig2}(a) as the unit quantum circuit for
measuring the depth of circuits constructed by the scheme. In addition, 
let $\varepsilon$ denote uniform original bias for qubits. We assume that 
$0\le\varepsilon\le1$ because a negative original bias can be inverted 
in advance.

\begin{figure}
\includegraphics[scale=0.44]{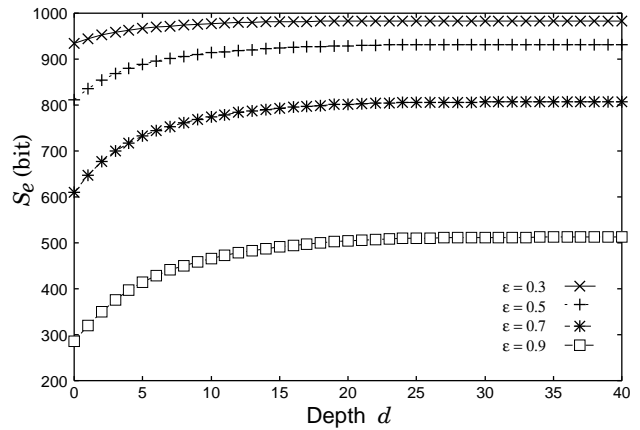}
\caption{\label{fig11}Plots of the effective entropy $S_e$ against
the depth of circuits from the simulation logs for $n=1000$.
The simulation started with qubits with uniform biases. The data points were
obtained for the different values of original bias: $\varepsilon=0.3$, 
$\varepsilon=0.5$, $\varepsilon=0.7$, and $\varepsilon=0.9$.}
\end{figure}

The logs recorded during each run of the simulation show that although 
the von Neumann entropy $S = -\mathrm{Tr}(\rho\log_2\rho)$ is preserved,
the number of qubits initialized after the boosting operations is 
much less than $n-S$ because $S_e$ is increased in the early steps of 
the boosting procedure. This is well illustrated as the plots of $S_e$ 
against the depth of the circuits (denoted by $d$) for $1000$ qubits 
with originally uniform biases, as shown in Fig.\ \ref{fig11}. For an 
example, let us take $\varepsilon=0.7$ in the figure. Although the 
value of the von Neumann entropy is $609.8$, the value of $S_e$ at 
$d=40$ is $806.8$. The number of qubits that can be initialized is 
reduced from $390$ to $193$ approximately.

\begin{figure}[bt]
\includegraphics[scale=0.6]{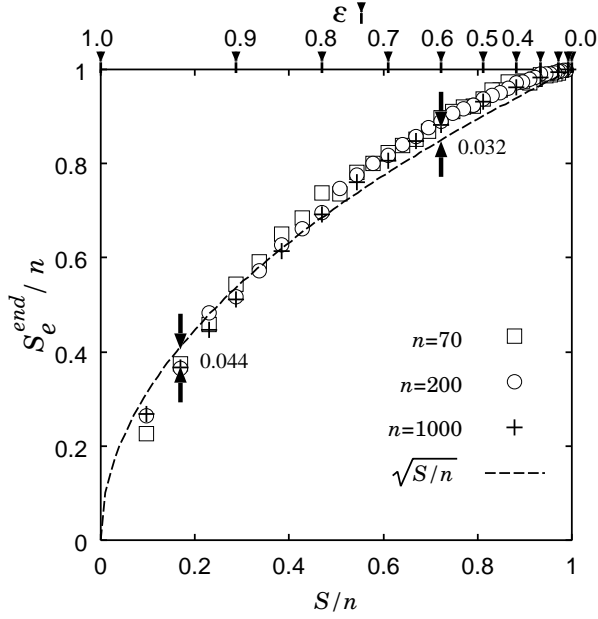}
\caption{\label{fig12}Plots of $S_e^{end}/n$ against $S/n$ for the 
different values of $n$: $n=70$, $n=200$, and $n=1000$. The values of
original uniform bias $\varepsilon$ are also shown on the horizontal 
axis. Each run of the simulation was terminated at the depth of 
$d=100$. The plots are data points from the simulation, all of which 
approximately lie on the curve of $S_e^{end}/n=\sqrt{S/n}$. The facing 
arrow pairs indicate the most distant data points from the curve in the 
vertical direction for $n=1000$, accompanied by the distance values 
(both upper and lower side).}
\end{figure}

\begin{figure}
\begin{center}
 \includegraphics[scale=0.64]{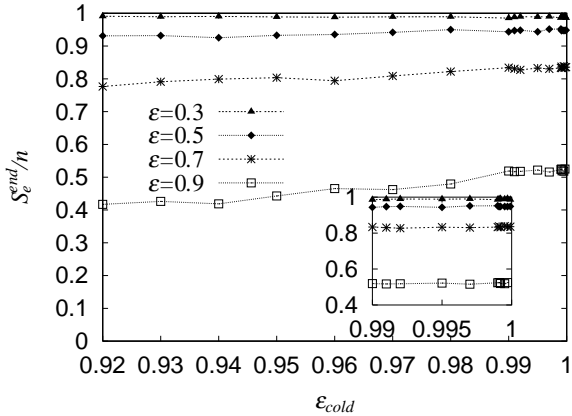}
\caption{\label{fig13}Plots of $S_e^{end}/n$ against $\varepsilon_{cold}$
 for several different values of original uniform bias $\varepsilon$.
 $n$ was set to $100$ for the data points in this figure.}
\end{center}
\end{figure}

The relation of $S_e^{end}$ (the terminal value of $S_e$) to $S$ was 
found to be of interest. We plot $S_e^{end}/n$ against $S/n$ for several
different values of $n$ in Fig.\ \ref{fig12}. The original biases 
were set to be uniform and the data were taken for 
$0\le\varepsilon\le0.975$. It suggests
\begin{equation}\label{eq31}
S_e^{end}/n\simeq \sqrt{S/n}
\end{equation}
for $n\ge 70$. In addition to this relation, it suggests that for 
$n\ge 70$ and $0\le\varepsilon\le 0.65$,
\begin{equation}\label{ineq32}
 S_e^{end}/n > \sqrt{S/n},
\end{equation}
in consequence of the fact that a value of $S_e^{end}/n$ was found to
be insensitive to a value of $n\ge70$ for this range of $\varepsilon$. 
The figure also shows that the data points of $S_e^{end}/n$ for 
$n=1000$ is slightly closer to the curve of $\sqrt{S/n}$ than those for 
smaller values of $n$. Furthermore, more than $60\%$ of the increase in 
$S_e$ occurs during $d \le 5$ when $n=1000$; the number of qubits to 
which a qubit is classically correlated is nearly $3^5=243<1000$ for 
this depth. Hence the growth in $S_e/n$ during $d \le 5$ is almost 
unchanged for larger values of $n$. Thus, as $n$ becomes large, 
$S_e^{end}/n$ varies only slightly, getting closer to $\sqrt{S/n}$. 
Therefore, Eq.\ (\ref{eq31}) for $0\le\varepsilon\le 0.975$ and 
inequality\ (\ref{ineq32}) for $0\le\varepsilon\le 0.65$ are true for the 
circuits with larger widths, {\em i.e.}, they are true for 
${}^\forall n \ge 70$. Because the data points of $S_e^{end}/n$ are in 
the interval of $[\sqrt{S/n}-0.044,\sqrt{S/n}+0.032]$ when $n=1000$, 
the relation of $S_e^{end}$ to $S$ for $n\ge 1000$ and 
$0\le\varepsilon\le 0.975$ is given by
\begin{equation}\label{eq33}
 S_e^{end}= n\left(\sqrt{S/n} + \delta\right),
\end{equation}
where $\delta$ is a real number ($-0.05 < \delta < 0.04$).
This phenomenon is originated from the fact that the basic circuits
increase the effective entropy of 3 qubits input with independent
uniform biases unless their value is $0$ or $\pm 1$ as we have already
seen in Fig.\ \ref{fig5}. 
In addition, the value of $S_e^{end}/n$ is not sensitive to the preset 
value of $\varepsilon_{cold}$ as long as $\varepsilon_{cold}\ge0.92$ for 
$\varepsilon\lessapprox0.5$ and $\varepsilon_{cold}\ge0.99$ for 
$\varepsilon\gtrapprox0.5$ as shown in Fig.\ \ref{fig13}. Therefore, the
above relations, inequality\ (\ref{ineq32}) and Eq.\ (\ref{eq33}), are 
unchanged even if one presets $\varepsilon_{cold}$ in another way at 
one's option as long as $\varepsilon_{cold}$ is suitably large.

\begin{figure}[tb]
\includegraphics[scale=0.64]{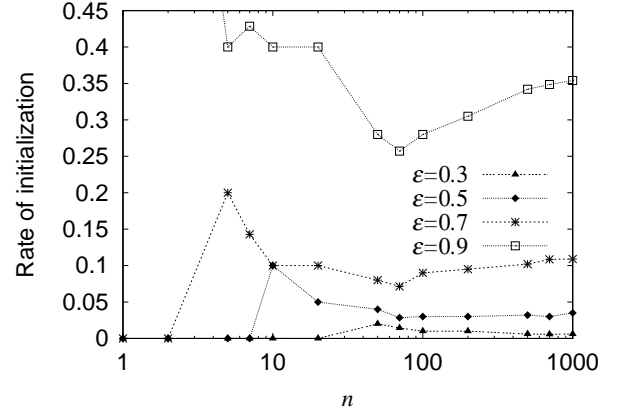}
\caption{\label{fig14}The initialization rate $l/n$ against $n$ for the 
values of uniform original bias: $\varepsilon=0.3$, $\varepsilon=0.5$, 
$\varepsilon=0.7$, and $\varepsilon=0.9$. $l$ is the number of cold 
qubits picked up at the last step in the simulation. 
For each data point, we chose the largest of 5 samples of $l$ that 
were calculated with different random seeds.}
\end{figure}

\begin{figure}
\begin{center}
\includegraphics[scale=0.42]{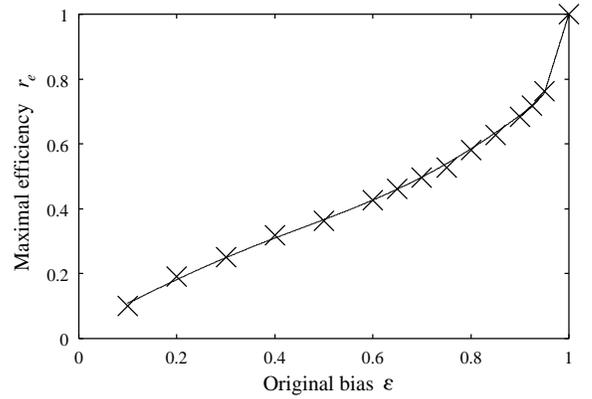}
\caption{\label{fig15}The maximal efficiency of initialization,
$r_e=(n-S_e^{end})/(n-S)$, for $n=1000$ against the uniform original 
bias $\varepsilon$. The crosses are data points from the simulation and 
the curve is the spline fit.}
\end{center}
\end{figure}

Moreover, the initialization rate $l/n$ is not rapidly improved by
increasing the value of $n$ as shown in Fig.\ \ref{fig14}, where $l$ is 
the number of cold qubits picked up at the end of the program. 
Although $l$ is expected to be close to $n-S_e^{end}$ for large values 
of $n$, only $l\simeq 0.7(n-S_e^{end})$ is achieved even when 
$n = 1000$ and $\varepsilon =0.9$. This is due to the increase in the 
threshold of the bias for cold qubits picked up at the terminal step, 
since it becomes large for large $n$ in order to avoid reduction in the 
population of the component state $|0_1\cdots0_l\rangle$.
There are still some other plots of data:
The maximal efficiency of initialization, $r_e=(n-S_e^{end})/(n-S)$, 
is plotted in Fig.\ \ref{fig15}. It is not large enough unless the 
original bias is almost $1$ if we regard the scheme as a classical 
data compression method.
Matrix calculations were conducted to verify the behavior of 
the automatically generated circuits for $n\le 13$, indicating the
reliability of our molecular simulation. Two of them have been already 
shown in the examples of Sec.\ \ref{example2}. The reliability of the 
simulation is also statistically evaluated in Appendix \ref{app1}.
The relation of $S_e^{end}$ to $S$ was obtained also for the biases that
are originally nonuniform when $n=70$, showing almost the same amount 
of increase in $S_e^{end}$ as that for originally uniform biases 
(see Appendix \ref{app2}).

One can conclude from the results of our molecular simulation, 
especially from Eq.\ (\ref{eq33}), that the number of initialized 
qubits after the application of the boosting scheme is
\begin{equation}
 l \le n\left(1-\sqrt{S/n}-\delta+\beta\right)
\end{equation}
for $n\ge 1000$ and $0\le\varepsilon\le 0.975$, where $S$ is the von 
Neumann entropy of the system, $\delta\in(-0.05,0.04)$, and $\beta$ is 
a small number that has appeared in inequality\ (\ref{eqlupper}) of 
Sec.\ \ref{sec3}. $\beta$ is negligible for large $l$, in particular 
for $n\ge1000$ and $\varepsilon\gtrapprox0.3$. When it is required that 
the bias of an initialized qubit picked up at the terminal step tends 
to $1$ with increasing $n$, $\beta=o(1)$. In addition, combining one of 
the results inequality\ (\ref{ineq32}) with inequality\ (\ref{eqlupper}), we find 
that, for $n\ge 70$ and $0\le\varepsilon\le 0.65$,
\begin{equation}
 l < n\left(1-\sqrt{S/n}+\beta\right).
\end{equation}

\section{Discussion}
Much work has been done on the theory of quantum data compression of mixed 
states \cite{SC95,HO97,MH98,JH98,HO99,NC2000,KS01}. Nevertheless, our 
numerical analysis is the first evaluation of large quantum circuits 
actually generated by a compression scheme for ensemble quantum 
computing as far as the authors know. Calculating an actual compression 
rate for a specific case is quite another subject than a theoretical 
proof of a limit to a compression rate. For an initialization of 
ensemble computers, the string to be compressed is a macroscopic string 
which is a probabilistic mixture of microscopic strings. It is probably 
impossible to generate a quantum circuit providing the optimal 
compression rate for this case as long as only the biases of qubits are 
referred to at each step of the circuit designing. It would be possible 
to construct the optimal compression circuit when it is generated by 
referring to the populations of component states. So far, efficient and 
effective initialization schemes are those which target the typical 
sequences \cite{NC2000,KKN03}.

The general difficulty in the evaluation of a quantum data compression 
scheme is caused by correlations among qubits after multiple-qubit 
operations. Our program has demonstrated that it does not cost so much 
resources to simulate a scheme for initial thermal states in which only 
permutations of component states are used since only classical 
correlations can be induced. The boosting scheme is one of this
special type of compression methods although its initialization rate is 
not large. There are many polynomial-time classical data compression 
schemes offering the average code length approximately equal to the 
Shannon entropy of the original probability distribution for suitably 
large $n$, {\em e.g.}, enumerative coding \cite{TMC73}, some of which 
were reconstructed as quantum algorithms \cite{CD96,BFGL98,CM00,JL02}. 
A common efficient classical data compression may be used for a more 
effective initialization as long as the number of required clean qubits 
of workspace is $\log_2[O(\mathrm{poly}(n))]$ \cite{KKN03}. With a 
proper molecular simulation, a quantum circuit constructed by a 
classical data compression scheme would be easily evaluated. 

To realize a high initialization rate, the value of the total binary entropy
$S_e$ at the terminal step in an initialization scheme must be nearly 
equal to that of the von Neumann entropy of the system. In the boosting 
scheme, as illustrated in Fig.\ \ref{fig11}, although the growth of $S_e$
is a little suppressed, the suppression is not sufficiently strong. The 
amount of correlation $S_e-S$ grows step by step, mainly in the early 
steps, in the bias boosting process. In other words, a classical string 
whose macroscopic bits are strongly correlated to each other can only be
compressed by the scheme without an increase in the amount of
correlation, although such a string does not exist initially in a common
NMR system. An effective initialization scheme must have an ability to 
suppress the growth in the total correlation.

Finally, we have to discuss the curious relation of $S_e^{end}$ to $S$ 
that has been suggested by the simulation results. The relation given by
Eq.\ (\ref{eq31}) is simple despite the complicated process of the 
growth in $S_e$. It is difficult to predict the growth without the 
molecular simulation. Indeed, there are nothing but a few trivial things
that support an intuitive interpretation of the relation:\\
1. $S_e^{end} = S$ when $\varepsilon = 0$ or $\varepsilon = 1$.\\
2. $S_e^{end}$ increases monotonically as $S$ increases, at least 
approximately.\\
But there might be also a profound physics in the simple relation, 
although it is, of course, a result for a special scheme.
We have calculated the growth of $S_e$ only in the circuits that
comprise the basic circuits shown in Fig.\ \ref{fig2}. It is of interest
to examine a growth in the sum of binary entropies in a similar scheme
with another basic circuits as well as that in other compression schemes.

\section{Conclusion}
A molecular simulation has been demonstrated to evaluate the
effectiveness of Schulman and Vazirani's boosting scheme for the 
initialization of NMR quantum computers. We have confirmed that a 
generated quantum circuit will enhance the biases of target qubits as 
it is designed to. However, our results have also indicated that the 
scheme is rather inefficient with respect to the initialization rate even 
for a large number of qubits. When we use 3-qubit basic circuits,
the rate is at most approximately $1-\sqrt{S/n}$ even for $n\ge 10^3$ 
as long as $0\le\varepsilon\le 0.975$,
where $S$ is the von Neumann entropy of a whole $n$-qubit 
ensemble. 
This is owing to a large increase in the sum of the binary entropies of 
individual qubits. This increase means that a large amount of classical 
correlation is induced in the macroscopic $n$-qubit string. It is to be 
hoped that an advanced algorithm that suppresses the growth in the 
amount of correlation will be constructed to improve the rate.

\begin{acknowledgments}
The authors are grateful to the anonymous referees for their thoughtful
 comments. This work is supported by a grant termed CREST from Japan 
Science and Technology Agency.
\end{acknowledgments}

\appendix
\section{Reliability of the simulation}\label{app1}
Because our simulation uses a limited number of virtual molecules, 
there are, to some extent, errors in the output data. The errors 
comprise both systematic and statistical errors, since the time 
evolution of the virtual molecular system is deterministic in 
Program\ 1 while the setup of the system is dependent on the random 
seed (see also Appendix\ \ref{app3}). Here, we present a statistical 
evaluation of the reliability of the simulation system although the 
errors are expected to be small in the data used in Sec.\ \ref{sec7} 
as we have set the number of molecules suitably large.

\begin{figure}[hbpt]
\begin{center}
\includegraphics[scale=0.4]{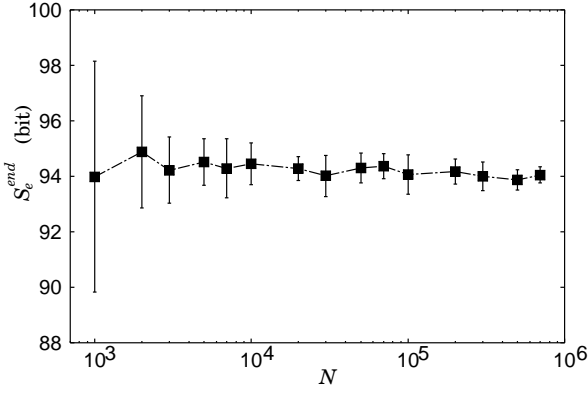}\caption{
\label{fig16} The average values of $S_e^{end}$ (solid squares) against 
the number of virtual molecules $N$ when $\varepsilon=0.5$ and $n=100$.
The error bars represent 99\% confidence intervals associated with each 
mean.}
\end{center}
\end{figure}

\begin{figure}[hbpt]
\begin{center}
\includegraphics[scale=0.4]{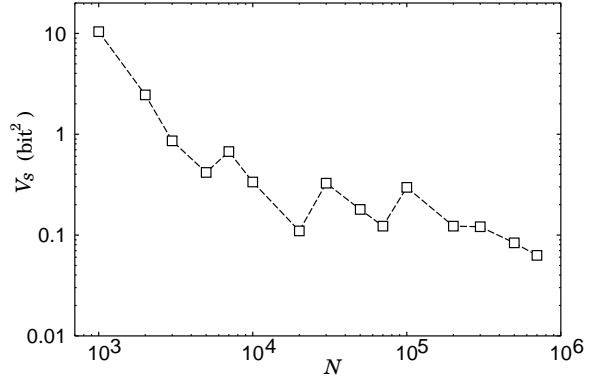}\caption{
\label{fig17} Plots of $V_s$ against the number of virtual molecules $N$. 
$V_s$ is the variance of the sample values of $S_e^{end}$ at each value
of $N$. The sample values are the same as those used in Fig.\ \ref{fig16}.
}
\end{center}
\end{figure}

In order to evaluate the reliability, the average of sample values
of $S_e^{end}$ was plotted against the number of molecules, $N$, as shown
in Fig.\ \ref{fig16}. $S_e^{end}$ is the value of the effective entropy
at the depth of $d = 100$ as no further increase in $S_e$ was found at 
larger depth. Considering the running time of the simulation, we chose 
the original uniform bias $\varepsilon=0.5$ and the number of qubits, 
$n=100$. The data were obtained by executing the program 60 times for 
each value of $N$ with different random seeds. In the figure, the error 
bars represent 99\% confidence intervals associated with each mean. 
The half width of the confidence interval is smaller than $0.4$ bit for the 
values of $N\ge 5.0\times 10^5$. Because the mean value of 
$S_e^{end}$ was found to be insensitive to the value of $N$ and we used 
the same number of samples for each data point, this indicates that the 
systematic errors are negligibly small for such large values of $N$. 
Along with the growth of $N$, the errors are reduced even further. 
Moreover, the variance of the samples, $V_s$, is less than 
$0.1 ~\mathrm{bit}^2$ when $N\ge 5.0\times 10^5$ (Fig. \ref{fig17}). 
Each value of $V_s$ was calculated with the 60 samples of $S_e^{end}$ 
at the corresponding value of $N$. This result shows that we do not need
to employ a statistical average value of $S_e^{end}$, {\em i.e.,} a 
single sample value of $S_e^{end}$ is reliable enough, when $N$ is set 
to such a large number. The non-averaged sample values of $S_e^{end}$ 
used in Sec.\ \ref{sec7} are reliable since we have set 
$N=5.0\times 10^6$.

\section{Initialization for nonuniform original biases}\label{app2}
We believe that the initialization for qubits with nonuniform
biases is as effective as that for qubits with uniform biases, in the 
boosting scheme. Consider $n$ qubits in the structure 
$\cdots\text{-A-B-A-B-A-B-}\cdots$ where $n$ is even; A and B represent 
the atom with original bias $\varepsilon_A\ge 0$ and that with
$\varepsilon_B\ge 0$, respectively. The initialization for this structure
is equivalent to the one for the structure 
$\text{A-A-}\cdots\text{-A-A-B-B-}\cdots\text{-B-B}$
since we arrange the pointers to the qubits in the order of the biases.

When $\varepsilon_A\gg \varepsilon_B$ ({\em i.e.}, $0\le\varepsilon_B\ll 1$), 
a pointer to A is rarely grouped with a pointer to B in early time steps
in the scheme; the increase in the effective entropy $S_{e,AB}$ occurs 
separately in $\text{A-A-}\cdots\text{-A-A}$ and in $\text{B-B-}\cdots\text{-B-B}$.
Together with Eq.\ (\ref{eq31}), this leads to that the terminal value
of $S_{e,AB}$ is given by 
\begin{equation}
 S_{e,AB}^{end}\simeq \frac{n}{2}\sqrt{2 S^{orig}_{A\cdots A}/n} 
+ \frac{n}{2}\sqrt{2 S^{orig}_{B\cdots B}/n},
\end{equation}
where $S^{orig}_{A\cdots A}\le n/2$ and  $S^{orig}_{B\cdots B}\le n/2$ 
are the von Neumann entropies in the original distribution for 
$\text{A-}\cdots\text{-A}$ and $\text{B-}\cdots\text{-B}$, respectively. 
The increase in the effective entropy is smaller than or approximately 
equal to that for qubits with originally uniform biases, since
\begin{equation}
S_{e,AB}^{end}/n\simeq \frac{1}{\sqrt{2n}}
\left(\sqrt{S^{orig}_{A\cdots A}}+\sqrt{S^{orig}_{B\cdots B}}\right)\le\sqrt{S/n},
\end{equation}
where $S=S^{orig}_{A\cdots A}+S^{orig}_{B\cdots B}$. On the other hand, 
when $\varepsilon_A\simeq\varepsilon_B$, the increase in the effective 
entropy is almost the same as that calculated for the originally uniform
biases. In this case,
\begin{equation}\label{eqB3}
 S_{e,AB}^{end}\simeq \sqrt{nS}.
\end{equation}

\begin{figure}[hbpt]
\begin{center}
\includegraphics[scale=0.68]{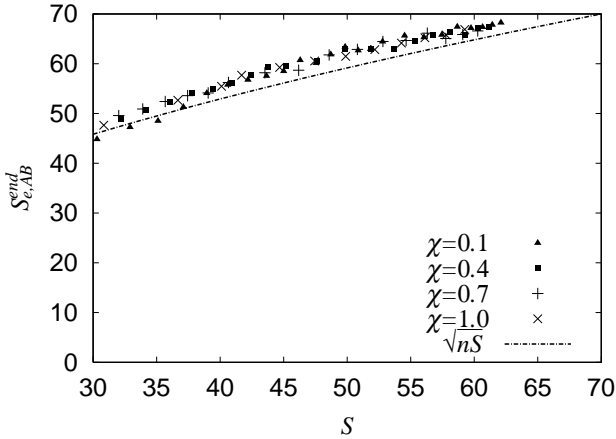}
\caption{\label{fig18} Plots of $S_{e,AB}^{end}$ against $S$ for 
$\chi=0.1$, $\chi=0.4$, $\chi=0.7$, and $\chi=1$ when $n=70$.}
\end{center}
\end{figure}

Now we utilize the coefficient $\chi$ and set 
$\varepsilon_B=\chi\varepsilon_A$ to calculate the increase in the 
effective entropy in the scheme when $n=70$ for several different values
of $\chi$ ($0<\chi\le 1$). The relation of $S_{e,AB}^{end}$ to $S$ is well
depicted in Fig.\ \ref{fig18}. The figure shows that, as long as
$0.1\le\chi\le 1$, $S_{e,AB}^{end}$ is not sensitive to the value of 
$\chi$ and all the points approximately lie on the curve of 
$S_{e,AB}^{end}=\sqrt{nS}$. Hence, the use of Eq.\ (\ref{eqB3}) is 
justified for $n=70$.
One can conclude that the number of initialized qubits is 
\begin{equation}
 l_{AB} \lessapprox n\left(1-\sqrt{S/n}\right)
\end{equation} 
for any realistic value of $\chi$ ($0.1\le\chi\le 1$). Although the 
relation has been verified in the structure that comprises two species 
of qubits, similar results are expected to be given for molecules that 
consist of atoms with many different original biases.

\section{Details of the simulation algorithm}\label{app3}
Here, we show the details of the simulation algorithm for the bias 
boosting scheme in the following program of Program\ 1, which has 
been explained in Sec.\ \ref{sec5}. 
  \\\vspace{8pt}\par\noindent
PROGRAM\ 1: The simulation algorithm for generating an initialization
circuit based on the boosting scheme. 
Here, $\text{frand()}$ is a function that returns a pseudo random real 
number between $0$ and $1$. A qubit with a bias greater than 
$\varepsilon_{cold}$ is regarded as a cold qubit. 
\\ \\
\textbf{procedure GEN\_INIT\_CIRCUIT:}\\
\textbf{begin}\\
\verb|   | \textbf{call} \textbf{CREATE\_SYSTEM}\\
\verb|   | Create tbl[ ], a table of pointers to individual qubits\\ 
\verb|   | (tbl[1] is the uppermost end).\\
\verb|   | Arrange tbl[ ] so that the pointer to a qubit with a\\
\verb|   | larger \hspace{1pt}bias \hspace{1pt}appears \hspace{1pt}upper \hspace{1pt}side \hspace{1pt}and \hspace{1pt}that \hspace{1pt}with \hspace{1pt}a\\
\verb|   | smaller appears lower side.\\
\verb|   | \textbf{while} new cold qubits are obtained in last $s_t$ steps\\
\verb|   | \verb|      |\textbf{and} depth of output circuit $\le$ preset max-\\
\verb|   | \verb|      |imum depth; \textbf{do}\\
\verb|   | \verb|   | $j$ $\leftarrow$ minimum $j$ such that
$\varepsilon_{\text{tbl[}j\text{]}}$ $\le$ $\varepsilon_{cold}$\\
\verb|   | \verb|   | \textbf{while} $j$ $\le$ $n$ - 2 \textbf{do}\\
\verb|   | \verb|   | \verb|   | \textbf{call} \textbf{BOOST}(tbl[$j$], tbl[$j+1$], tbl[$j+2$])\\
\verb|   | \verb|   | \verb|   | \textbf{if} $\varepsilon_{\text{tbl[}j\text{]}}$ is increased \textbf{then}\\
\verb|   | \verb|   | \verb|   | \verb|   | Output the circuit applied in the last\\
\verb|   | \verb|   | \verb|   | \verb|   | call of \textbf{BOOST} procedure with the\\
\verb|   | \verb|   | \verb|   | \verb|   | indication of handled qubits, tbl[$j$],\\
\verb|   | \verb|   | \verb|   | \verb|   | tbl[$j+1$], and tbl[$j+2$].\\
\verb|   | \verb|   | \verb|   | \textbf{else}\\
\verb|   | \verb|   | \verb|   | \verb|   | Undo the last \textbf{BOOST}.\\
\verb|   | \verb|   | \verb|   | \textbf{endif}\\
\verb|   | \verb|   | \verb|   | $j$ $\leftarrow$ $j$ + 3\\
\verb|   | \verb|   | \textbf{endwhile}\\
\verb|   | \verb|   | Re-arrange tbl[ ].\\
\verb|   | \verb|   | Take a log of tbl[ ] and $\varepsilon_{\text{tbl[ ]}}$ for simulation use.\\
\verb|   | \textbf{endwhile}\\
\textbf{end}\\ \\
\textbf{procedure CREATE\_SYSTEM:}\\
Create $N$ molecules. (A molecule consists of $n$ binary digits.)\\
\textbf{begin}\\
\verb|   | \textbf{for} $m$ $\leftarrow$ 1 \textbf{to} $N$ \textbf{do}\\
\verb|   | \verb|   | \textbf{for} $i$ $\leftarrow$ 1 \textbf{to} $n$ \textbf{do}\\
\verb|   | \verb|   | \verb|   | \textbf{if} frand() $<$ (1 + $\varepsilon_i$) / 2 \textbf{then}\\
\verb|   | \verb|   | \verb|   | \verb|   | $i$th bit of $m$th molecule $\leftarrow$ 0\\
\verb|   | \verb|   | \verb|   | \textbf{else}\\
\verb|   | \verb|   | \verb|   | \verb|   | $i$th bit of $m$th molecule $\leftarrow$ 1\\
\verb|   | \verb|   | \verb|   | \textbf{endif}\\
\verb|   | \verb|   | \textbf{endfor}\\
\verb|   | \textbf{endfor}\\
\textbf{end}\\ \\
\textbf{procedure BOOST}($a$, $b$, $c$)\textbf{:}\\
\textbf{begin}\\
\verb|   | Consider the trio, $a$th bit, $b$th bit, and $c$th bit.\\
\verb|   | \textbf{for} $m$ $\leftarrow$ 1 \textbf{to} $N$ \textbf{do}\\
\verb|   | \verb|   | Apply Fig.\ \ref{fig2}(a) to the trio on $m$th molecule.\\
\verb|   | \textbf{endfor}\\
\verb|   | \textbf{call} \textbf{UPDATE\_BIAS}($a$)\\
\verb|   | \textbf{call} \textbf{UPDATE\_BIAS}($b$)\\
\verb|   | \textbf{call} \textbf{UPDATE\_BIAS}($c$)\\
\verb|   | \textbf{if} $\varepsilon_{b}$ $<$ 0 \textbf{then}\\
\verb|   | \verb|   | \textbf{for} $m$ $\leftarrow$ 1 \textbf{to} $N$ \textbf{do}\\
\verb|   | \verb|   | \verb|   | Apply Fig.\ \ref{fig2}(b) to the trio on $m$th\\
\verb|   | \verb|   | \verb|   | molecule.\\
\verb|   | \verb|   | \textbf{endfor}\\
\verb|   | \verb|   | \textbf{call} \textbf{UPDATE\_BIAS}($b$)\\
\verb|   | \textbf{endif}\\
\textbf{end}\\ \\
\textbf{procedure UPDATE\_BIAS}($i$)\textbf{:}\\
\textbf{begin}\\
\verb|   | $s$ $\leftarrow$ 0\\
\verb|   | \textbf{for} $m$ $\leftarrow$ 1 \textbf{to} $N$ \textbf{do}\\
\verb|   | \verb|   | \textbf{if} $i$th bit of $m$th molecule is 0 \textbf{then}\\
\verb|   | \verb|   | \verb|   | $s$ $\leftarrow$ $s$ + 1\\
\verb|   | \verb|   | \textbf{endif}\\
\verb|   | \textbf{endfor}\\
\verb|   | $p$ $\leftarrow$ $s$ / $N$\\
\verb|   | $\varepsilon_i$ $\leftarrow$ 2 * $p$ - 1\\
\textbf{end}


\bibliography{letter-0}

\end{document}